\documentclass[prl,onecolumn,showpacs,superscriptaddress,floatfix]{revtex4}
\usepackage{graphicx}

\begin{document}
\title{Impact of weak localization on wave dynamics: Crossover from quasi-1D to slab geometry}

\author{Z.Q.~Zhang}
\affiliation{Department of Physics, Hong Kong University of Science and Technology, Clear Water Bay, Kowloon, Hong Kong}

\author{S.K.~Cheung}
\affiliation{Department of Physics, Hong Kong University of Science and Technology, Clear Water Bay, Kowloon, Hong Kong}

\author{X.~Zhang}
\affiliation{Department of Physics, Hong Kong University of Science and Technology, Clear Water Bay, Kowloon, Hong Kong}

\author{A.A.~Chabanov}
\affiliation{Department of Physics, Queens College of the City University of New York, Flushing, New York 11367, USA}

\author{A.Z.~Genack}
\affiliation{Department of Physics, Queens College of the City University of New York, Flushing, New York 11367, USA}

\date{\today}

\begin{abstract}
We study the dynamics of wave propagation in nominally diffusive samples by solving the Bethe-Salpeter equation with recurrent
scattering included in a frequency-dependent vertex function, which renormalizes the mean free path of the system. We calculate
the renormalized time-dependent diffusion coefficient, $D(t)$, following pulsed excitation of the system. For cylindrical
samples with reflecting side walls and open ends, we observe a crossover in dynamics in the transformation from a quasi-1D to a
slab geometry implemented by varying the ratio of the radius, $R$, to the length, L. Immediately after the peak of the
transmitted pulse, $D(t)$ falls linearly with a nonuniversal slope that approaches an asymptotic value for $R/L\gg 1$. The value
of $D(t)$ extrapolated to $t=0$, depends only upon the dimensionless conductance $g$ for $R/L \ll 1$ and upon $kl_0$ and $L$ for
$R/L \gg 1$, where $k$ is the wave vector and $l_0$ is the bare mean free path.
\end{abstract}

\maketitle

\begin{section}{I. Introduction}
Weak localization (WL) of electronic and classical waves arises from the interference of counterpropagating partial waves in
closed loops. Its impact upon electronic conductance \cite{Mesobook} has been widely studied using steady-state methods such as
magnetoresistance \cite{Bergmann}, and can be directly visualized as enhanced retroreflection of light from random samples
\cite{CBS}. Partial waves with trajectories over a wide range of lengths contribute to these phenomena. Most of the studies in
the past were focused on the static transport properties.  For instance, for a bulk system, WL makes all states localized in one
and two dimensions.  In three dimensions, it renormalizes the diffusion constant and can lead to the Anderson localization
transition when the Ioffe-Regel criterion $k\ell<1$ is met, where $k$ is the wave vector and $\ell$ is the mean free path
\cite{Sheng}. It is of great interest to investigate the variation of WL in the time domain
\cite{Altshuler,Berkovits90,Weaver94,Mirlin00,Chabanov02}.  Since  the impact of WL  increases with pathlength, which is
proportional to time, it can dictate the dynamical behavior of wave propagation in a finite-sized sample. Though it has not been
practical to isolate paths of specific lengths in studies of electronic conductance, this can be accomplished for classical
waves in time-resolved measurements of pulsed transmission through random media
\cite{Chabanov02,McCall87,GenDrake89,Alfano90,Lagendijk97,Zang99}. Time-resolved transmission measurements
\cite{McCall87,GenDrake89} have generally been consistent with diffusion theory, which predicts a simple exponential decay of
the average transmission.  The asymptotic decay rate due to leakage from the sample is $1/t_{D}=\pi^{2}D_{0}/(L+2z_{0})^{2}$,
where $t_{D}$ is the diffusion time, $D_{0}$ is the diffusion coefficient, $L$ is the sample thickness, and $z_{0}$ is the
extrapolation length. Recently, however, Chabanov \textit{el al.} \cite{Chabanov02} observed nonexponential decay of pulsed
microwave transmission in a quasi-1D sample which was characterized via a ``time-dependent diffusion coefficient," $D(t)$. In
fact, nonexponential behavior has been predicted for electronic systems.   It was predicted that, when time is much greater than
the Heisenberg time, $t_{H}=\pi^2gt_{D}$, where $g$ is the dimensionless conductance, the decay of the electron survival
probability follows a log-normal behavior due to the presence of prelocalized states arising from some rare configurations of
disorder in the medium \cite{Altshuler,Apalkov02}. Recently, in a supermatrix model calculation by Mirlin \cite{Mirlin00}, an
analytic expression has been found for the tail of the electron survival probability when $t \ll t_H$.  For the case of the
orthogonal ensemble, it can be expressed in term of a ``time-dependent diffusion coefficient," $D(t)/D_{0}=1-2t/t_H$. Although
such a linear decay of $D(t)$ was found in microwave experiments \cite{Chabanov02}; nevertheless, $D(t)$ did not extrapolate to
the bare diffusion coefficient at $t=0$ and the scaling behavior of the slope of $D(t)$ did not agree with the above expression.
Nonexponential decay of pulsed transmission has also been reported recently in numerical simulations in 2D \cite{Haney03} and in
a self-consistent diffusion theory, which includes recurrent scattering \cite{Bart04}.

In this work, we solve the Bethe-Salpeter equation with recurrent scattering included in a manner that satisfies the Ward
Identity \cite{Vollhardt80,Kirkpatrick85} to obtain the average time-dependent intensity transmitted through a random sample
following pulsed excitation. Recurrent scattering is treated in the framework of self-consistent localization theory
\cite{Vollhardt80,Kirkpatrick85} and is included in a frequency-dependent vertex function, which renormalizes the mean free path
of the system. We calculate the renormalized time-dependent diffusion coefficient, $D(t)$, following pulsed excitation. The
sample is cylindrical with reflecting side walls and open ends. By changing the ratio of the longitudinal dimension $L$ and the
radius $R$, a continuous transition can be made between two key experimental geometries: a quasi-1D geometry with $L\gg R$,
which is commonly employed in microwave experiments, and a slab with $L\ll R$, which is the typical optical geometry. For the
quasi-1D geometry, we find $D(t)/D_{0} = A -(2B/\pi^2gt_{D})t$ for $t\ll t_H$. The constant part, $A$, is universal, depending
only on $g$, while $B$ is nonuniversal and depends upon $L/\ell_{0}$, where $\ell_{0}$ is the bare mean free path, as well as
upon $g$. In the limit $g\gg 1$ and $L/\ell_{0}\gg 1$, our results coincide with supersymmetry calculations in Ref. [7]. For
moderate values of $g$, $g\ge 5$, our results are in agreement with experiment \cite{Chabanov02}. For the slab geometry, $D$
approaches a nearly constant renormalized value, being equal to $D_{0}(1-1.03/(k\ell_{0})^2)$ in the limit $L/\ell_{0}\gg 1$.
This is close to the result of WL theory for a bulk system.
\end{section}

\begin{section}{II. Theory}
We consider a plane wave pulse normally incident on the front surface of a random sample at $z = 0$. We assume that there is no
gain or absorption in the medium and that scattering is isotropic. The ensemble-averaged intensity $\langle
I(t,\mathbf{r})\rangle$ within the sample is obtained from the Fourier Transform of the frequency correlation function of the
scalar field $C_{\Omega}(\omega;\mathbf{r})=\langle\phi_{\Omega^{+}}(\mathbf{r}) \, \phi^{*}_{\Omega^{-}}(\mathbf{r})\rangle$,
i.e.,
\begin{equation}
\langle I(t,\mathbf{r})\rangle=\frac{1}{2\pi}\int d\omega exp(-i \omega t) C_{\Omega}(\omega;\mathbf{r}),
\end{equation}
where $\Omega^{\pm}=\Omega\pm\omega/2$, $\Omega$ is the central frequency, $\omega$  is the modulation frequency and
$\phi_{\Omega}(\mathbf{r})$ is the wave field at position $\mathbf{r}$ inside the sample.  In fact, it was Prof. George Papanicolaou who
pointed out the importance of this frequency correlation function in the study of wave dynamics in randomly layered media
\cite{George86,George87,George90}. $C_{\Omega}(\omega;\mathbf{r})=\langle\phi_{\Omega^{+}}(\mathbf{r}) \,
\phi^{*}_{\Omega^{-}}(\mathbf{r})\rangle$, is obtained from the space-frequency correlation function
$C_{\Omega}(\omega;\mathbf{r},\mathbf{r}')=\langle\phi_{\Omega^{+}}(\mathbf{r}) \, \phi^{*}_{\Omega^{-}}(\mathbf{r}')\rangle$, which
satisfies the following Bethe-Salpeter equation,
\begin{eqnarray}
C_{\Omega}(\omega;\mathbf{r},\mathbf{r}')\!\!\! &=&\!\!\!\langle\phi_{\Omega^{+}}(\mathbf{r})\rangle \,
\langle\phi^{*}_{\Omega^{-}}(\mathbf{r}')\rangle \nonumber \\ + \int\!\!\!\! &d\mathbf{r}_{1}&\!\!\! d\mathbf{r}_{2} \,
d\mathbf{r}_{3} \, d\mathbf{r}_{4} \langle G_{\Omega^{+}} (\mathbf{r},\mathbf{r}_{1})\rangle\langle \,
G_{\Omega^{-}}(\mathbf{r}',\mathbf{r}_{3})\rangle \, \nonumber \\ &\times& \!\!\! U_{\Omega}(\omega \,
;\mathbf{r}_{1},\mathbf{r}_{2} \, ;\mathbf{r}_{3},\mathbf{r}_{4}) \, C_{\Omega}(\omega ;\mathbf{r}_{2},\mathbf{r}_{4}) \, ,
\label{} \end{eqnarray}
where $\langle\phi_{\Omega}(\mathbf{r})\rangle$ is the coherent source inside the sample, and $\langle
G_{\Omega}(\mathbf{r},\mathbf{r}_{1}) \rangle = -{\exp(i\kappa|\mathbf{r}-\mathbf{r}_{1}|)\over4\pi|\mathbf{r}-\mathbf{r}_{1}|}$
is the ensemble-averaged Green's function that represents the coherent part of wave propagation from $\mathbf{r_1}$ to
$\mathbf{r}$. The complex wavevector $\kappa=k+{i\over 2\ell}$ describes the ballistic propagation inside the random media,
where $k={\Omega\over v}$ is the wavevector, $v$ is the phase velocity, and $\ell$ is the scattering mean free path, which is
determined from the imaginary part of the self-energy of $\langle G\rangle$. In the absence of weak localization, the bare mean
free path, $\ell_0$ , is determined from the single-scattering diagram via $\ell_0=1/n\sigma$, where $n$ is the density of
scatterers and $\sigma$ is the total scattering cross section. The vertex function $U_{\Omega}$ represents the sum of all
irreducible vertices. Here we approximate $U_{\Omega}$ as
\begin{eqnarray}
\lefteqn{U_{\Omega}(\omega;\mathbf{r}_{1},\mathbf{r}_{2};
\mathbf{r}_{3},\mathbf{r}_{4})=}\nonumber\\
& &{4\pi\over\ell_{0}} [1+\delta(\omega,k)]\, \delta(\mathbf{r}_{1}-\mathbf{r}_{2}) \, \delta(\mathbf{r}_{1}-\mathbf{r}_{3}) \,
\delta(\mathbf{r}_{1}-\mathbf{r}_{4}). \label{}
\end{eqnarray}
The first term in the vertex function with a scattering strength $4\pi/\ell_{0}$ represents self-avoiding paths and generates all the ladder
diagrams that give rise to wave diffusion when $L\gg \ell_0$ \cite{Lagendijk88}. It is worth mentioning that in the absence of the second
term Eq. (2) is equivalent to the radiative transport equation, which has been extensively used by Prof. George Papanicolaou in the study of
multiple scattering in random media \cite{George73}. The second term with a vertex strength $4\pi\delta(\omega,k)/\ell_0$ represents WL
contribution to the vertex function. The presence of this term renormalizes the bare mean free path to a frequency-dependent mean free path,
i.e., $\ell(\omega,k)=\ell_{0}/[1+\delta(\omega,k)]$. For flux conservation to hold, the Ward Identity \cite{Vollhardt80,Kirkpatrick85}
requires that the mean free path $\ell$ that appears in $\langle G\rangle$ should also be replaced by the same $\ell(\omega,k)$.  Since the
second term represents recurrent scatterings, it is obtained by summing all maximally-crossed diagrams due to weak localization in the tube
geometry. The assumption of point-like scattering for the WL contribution is justified as long as the system is far from the localization
threshold, so that the renormalized mean free path is scale independent.  In a bulk, the renormalization factor $\delta(\omega,k)$ can be
obtained from the renormalized diffusion coefficient \cite{Kirkpatrick85}, which can be written as

\begin{equation}
\frac{1}{D(\omega,k)}={\frac{1}{D_0}} [1+\frac{2\pi v}{k^2}\tilde{G}(\omega;\mathbf{r},\mathbf{r})], \label{}
\end{equation}
where $D(\omega,k)=v\ell(\omega,k)/3$, and $\tilde{G}$ is the Green's function that satisfies the diffusion equation in the frequency
domain:
\begin{equation}
(D_0\nabla^2+i\omega)\tilde{G}(\omega;\mathbf{r},\mathbf{r}')=-\delta(\mathbf{r}-\mathbf{r}'),
\end{equation}
The diagonal term of the Green's function,  $\tilde{G}(\omega;\mathbf{r},\mathbf{r}')$, represents the return probability of
waves that travel diffusively inside the sample.   For the geometry considered here, we solve for $\tilde{G}$  by using the
boundary conditions for a tube of radius $R$ and length $L$ , with perfectly reflecting cylindrical wall and open ends.  The
result can be written as
\begin{equation}
\tilde{G}=\tilde{G}_1+\tilde{G}_2,
\end{equation}
with
\begin{equation}
\tilde{G_1}(\omega;z=z')=\frac{2}{\pi R^2 \tilde{L}}\sum_{n=1}^{n_c}\frac{\sin^2[q_n(z+z_0)]}{-i\omega+D_0q_n^2}
\end{equation}
and
\begin{eqnarray}
\tilde{G_2}(\omega;z=z')=\frac{1}{2\pi D_0 \tilde{L}}\sum_{n=1}^{n_c}\sin^2[q_n(z+z_0)] \nonumber\\
\times\ln\left[\frac{-i\omega+D_0(q_n^2+\alpha^2/\ell_0^2)}{-i\omega+D_0(q_n^2+1/R^2)}\right],
\end{eqnarray}
where $q_n=n\pi/\tilde{L}$ is the transverse momentum, $n_c=\alpha\tilde{L}/\pi l_0$ is the upper momentum cutoff in the
z-direction, $\tilde{L}=L+2z_0$ is the effective length of the sample and $z_0=0.7104\ell_{0}$ is the extrapolation length
\cite{Sheng}. The factor $\alpha/\ell_{0}$ in Eq. (8) denotes the upper momentum cutoff in the transverse direction. $\alpha$ is
chosen as $1$  in our calculations unless otherwise specified. The $\tilde{G}_1$ term arising from all the diffusive modes that
are uniform in the transverse direction, i.e., $\vec{q}_{\perp}=0$ , is responsible for the continuous change of decay rate in
the ensemble-averaged transmitted intensity $\langle I(t)\rangle$. The existence of this term is due to the reflecting boundary
conditions at the cylindrical wall at which $\nabla\tilde{G}=0$ normal to the surface. The $\tilde{G}_2$ term represents the
contributions of the transverse modes other than $\vec{q}_{\perp}=0$. Here we have assumed that $\tilde{G}_2$ is independent of
the transverse position, which is valid when $R\gg\ell_0$. This term becomes important when $2R/L$ is not small. In the limit of
slab geometry, i.e., $2R/L\gg1$, $\tilde{G}_1$ is insignificant and $\tilde{G}_2$ becomes the main contribution. Eqs. (7) and
(8) indicate that the renormalized mean free path is z-dependent. In order to simplify our calculations, we take the spatial
average along the z-axis and replace the factor of $sin^{2}[q_{n}(z+z_{0})]$ by $1/2$. This simplification is valid as long as
$L \gg \ell_0$. The self-consistent theory of WL requires us to replace $D_0$ in Eq. (7) and (8) by $D(\omega,k)$
\cite{Vollhardt80,Kirkpatrick85}. Using Eqs. (3,4,6-8), we finally obtain

\begin{equation}
\delta(\omega,k)=\delta_{1}(\omega,k)+\delta_{2}(\omega,k),
\end{equation}
with
\begin{equation}
\delta_1(\omega,k)=\frac{v}{2N\tilde{L}} \sum^{n_c}_{n=1}\frac{1}{-i\omega+D(\omega,k) q_n^2} \label{}
\end{equation}
and
\begin{equation}
\delta_2(\omega,k)=\frac{3}{2k^2l_0\tilde{L}}\sum^{n_c}_{n=1} \ln\!\left(\frac{-i\omega+D(\omega,k) (q_n^2+\alpha^2/l_0^2)}
{-i\omega+D(\omega,k) (q_n^2+1/R^2)}\right), \label{}
\end{equation}
where $N=(kR)^2/4$ is the number of transverse modes in the tube. Eq. (11) requires $R>\ell_0/\alpha$, otherwise,
$\delta_2(\omega=0)<0$. In our calculations, we solve Eqs. (10) and (11) for $\delta(\omega,k)$ self-consistently with
$D(\omega,k)=v\ell(\omega,k)/3=v\ell_0/[3(1+\delta(\omega,k))]$. Physically, the self-consistency of $\ell(\omega,k)$ or
$D(\omega,k)$ assures the continuous renormalization due to recurrent scattering paths. In the static limit and when $L/\ell_0$
is large, it can be shown that $\delta_1(\omega=0)=1/(3g-1)$, where $g=4N\ell_0/3\tilde{L}$ is the dimensionless conductance.
This result implies that localization transition occurs at $g=1/3$ or $N=\tilde{L}/4\ell_0$ in quasi-1D geometry.

In our calculation, we assume both the excitation intensity and the scattered intensity are uniform over the transverse
cross-section of the tube. Eq.~(1) can now be written as \cite{Zang99}
\begin{eqnarray}
C_{\Omega}(\omega\!\!\!\!\!&,&\!\!\!\!z)=
\exp\!\left({i\omega z\over v}-{z\over\ell_{0}}\right) \nonumber \\
&+&\!\!\frac{1}{4\pi\ell(k,\omega)}\int_{0}^{L}\!\!dz' H(\omega,z-z')\, C_{\Omega}(\omega,z') \, ,
\end{eqnarray}
where
\begin{eqnarray}
H(\omega ,z-z')&=&\pi \int d\rho^{2} \nonumber \\
&\times&{\exp \!\left[ \left({i\omega\over v} -{1\over\ell(\omega,k)}\right)\!\sqrt{\rho^{2}+(z-z')^{2}}\right] \over
\rho^{2}+(z-z')^{2}}.\nonumber \\ \,
\end{eqnarray}
Eq. (12) is numerically solved for $C_{\Omega}(\omega,z)$. The transmitted intensity $\langle I(t,L)\rangle$  is then calculated
from the Fourier transform of $C_{\Omega}(\omega,L)$ , from which we obtain the renormalized time-dependent diffusion
coefficient by using the relation $D(t)/D_{0}=-\tau_{D}\, d\ln\langle I(t)\rangle/dt$, for $t>\tau_{D}$, where
$\tau_D=\tilde{L}^2/\pi^2D_0$ is the diffusion time \cite{Chabanov02}.
\end{section}

\begin{section}{III. Results and Discussions}
The time evolution of wave propagation is described by the renormalized diffusion coefficient $D(t)/D_0$, which, in turns,
depends on the behavior of $\delta(\omega)$. Thus it is important to understand the general behaviors of $\delta_1(\omega)$ and
$\delta_2(\omega)$. In Fig. 1, we show the functions $\delta_1(\omega)$ and $\delta_2(\omega)$ for a typical case of $k\ell_0=4$
and $L=20\ell_0$. For $\delta_1(\omega)$ we choose $R=3.5\ell_0$ ($2R/L=0.35$), which corresponds to $N=50$. For
$\delta_2(\omega)$ we choose $R$ as infinity, which corresponds to a slab geometry. The diffusion time in this case is
$\tau_D\cong139\ell_0/v$, which implies that the width of the frequency correlation function is, $\delta\omega\cong
2\pi/\tau_D=0.045$. From Fig. 1, a shape rise in the real part of $\delta_1(\omega)$ (curve a) is found when
$\omega<\delta\omega$.  This behavior suggests that significant renormalization of $D(t)$ will occur when $t>\tau_D$. In fact,
it is this sharp rise of $\delta_1$ at small $\omega$ that leads to the linear decay of $D(t)$ found previously
\cite{Chabanov02,Mirlin00}. However, the behavior of $\delta_2(\omega)$ is very different from that of $\delta_1(\omega)$. The
real part of $\delta_2(\omega)$ (curve c) increases slowly with decreasing $\omega$. This suggests a gradual renormalization of
$D(t)$ and will lead to a nearly constant $D(t)$ in the slab geometry. This is also consistent with the exponential decay of the
transmitted intensity found previously in the optical measurements \cite{Alfano90,Lagendijk97}.

In the inset of Fig. 1, we plot $\delta_1(0)$, $\delta_2(0)$ and $\delta(0)$ as functions of $2R/L$. It is clear that $\delta_1$
dominates when $2R/L\ll1$. As $2R/L$ is increased to $1$, $\delta_2$ becomes the dominant term. The saturation of $\delta_2$
when $2R/L>2$ implies that the behavior of $D(t)$ will not change much when $2R/L>2$.

\begin{figure}
\includegraphics [width=\columnwidth] {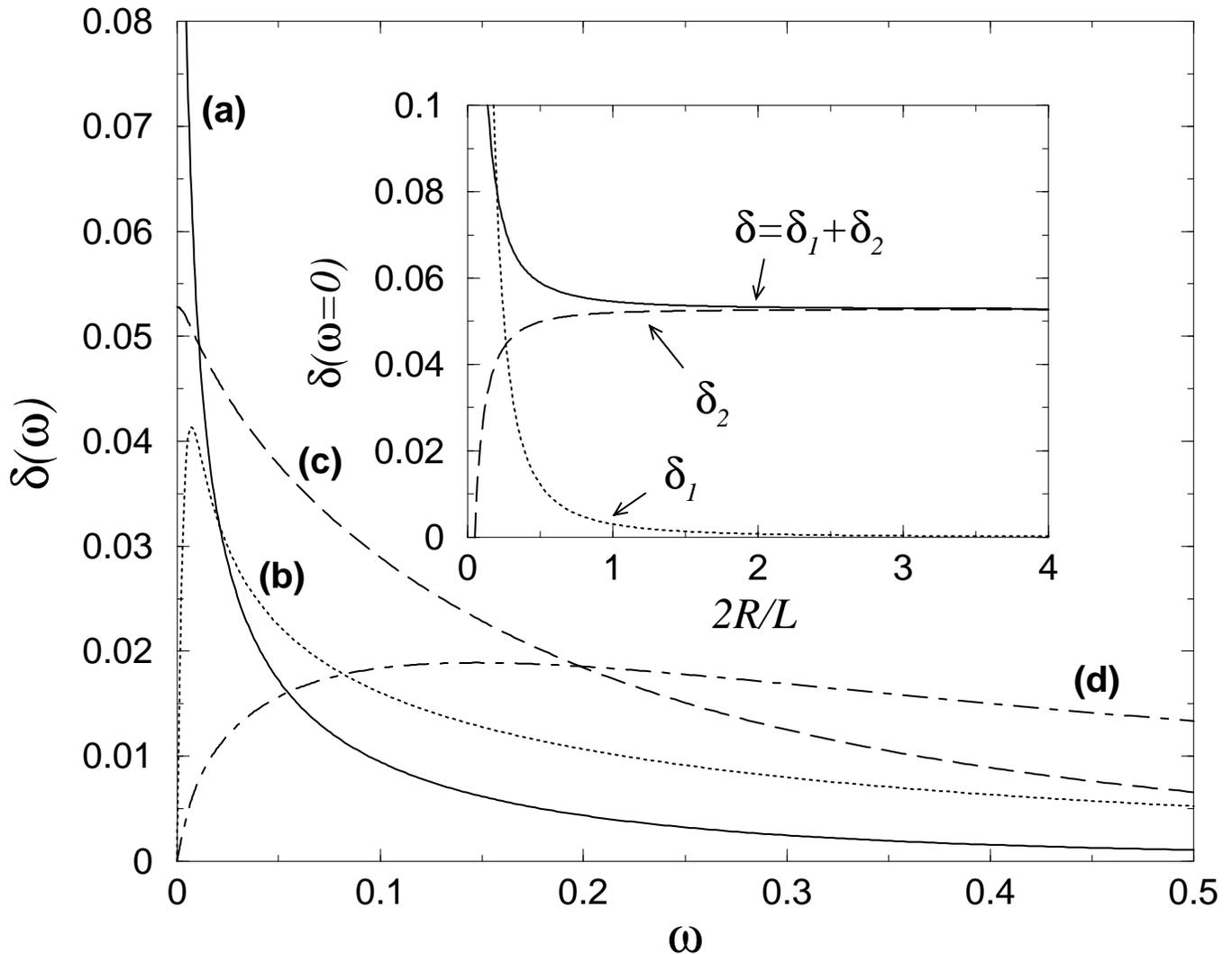}
\caption{$\delta(\omega)$ with $L=20\ell_0$ and $k\ell_0=4$ are plotted as functions of $\omega$. Here (a) is the real part and (b) is the
imaginary part of $\delta_1(\omega)$ when $R=3.5\ell_0$ ($N=50$), while (c) is the real part and (d) is the imaginary part of
$\delta_2(\omega)$ in the limit of $R\rightarrow\infty$ (slab geometry). In the inset: $\delta(\omega=0)$, $\delta_1(\omega=0)$ and
$\delta_2(\omega=0)$ for $L=20\ell_0$ and $k\ell_0=4$ are plotted as a function of $2R/L$.}
\end{figure}

\begin{subsection}{(a) quasi-1D system}
For quasi-1D systems, we first compare our theory with the microwave experiments reported in Ref. \cite{Chabanov02}. The results
are shown in Fig. 2 for $N=68$ and $L/\ell_{0}=6.4$ (sample A), $9.5$ (sample B), and $19.2$ (sample C), corresponding to
$\ell_{0}=9.5$ cm. The best fit is found at $\alpha=0.5$, and good agreement is found for Samples A and B. For Sample C, for
which $g=3$, however, our theory
predicts a smaller decay rate, reflecting the important role of localization effects, which are not included here. \\

\begin{figure}
\includegraphics [width=\columnwidth] {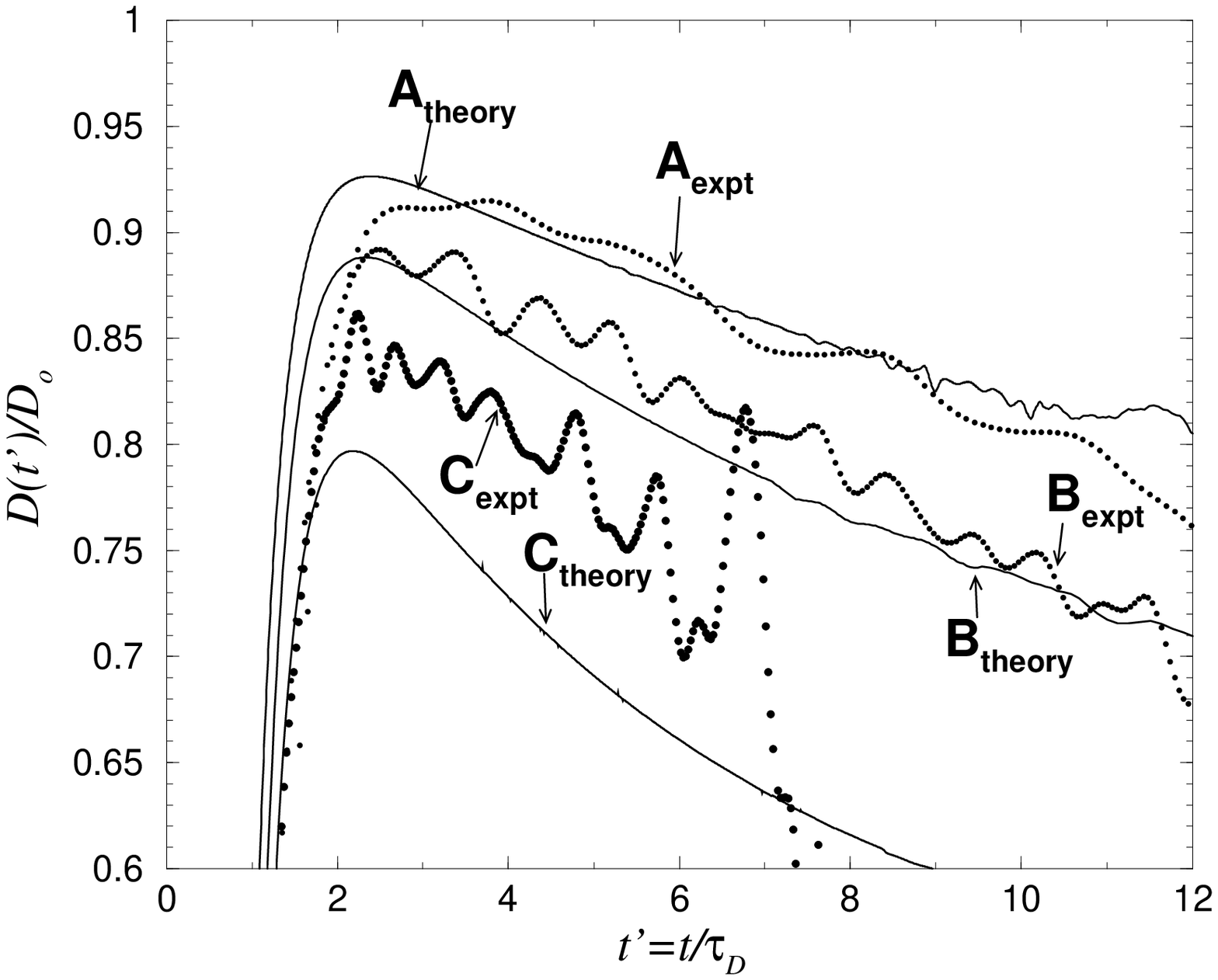}
\caption{$D(t)/D_0$ is plotted as a function of $t^\prime=t/\tau_D$ for three samples studied in Ref. \cite{Chabanov02}, with $N=68$,
$L/\ell_0=6.4$ (sample A), $9.5$ (sample B) and $19.2$ (sample C), where $D_0=42.5cm^2/ns$. The theoretical results for the corresponding
samples are shown in solid lines.}
\end{figure}

The linear region of the diffusion coefficient immediately after the peak can be fitted with $D(t)/D_0=A-Bt^{\prime\prime}$ ,
where $t''=2t/\pi^2g\tau_D$.  Mirlin's analytical result of Ref. \cite{Mirlin00} predicts $A=B=1$. Nevertheless, our
calculations \cite{Zhang04} show that $A$ is a universal function of $g$ and can be fitted by a single curve
$A=1.00-0.27/g-0.17/g^2$. In the limit of large $g$, $A$ approaches $1$.  However, the slope $B$ depends on both $g$ and
$L/\ell_0$. For each value of $L/\ell_0$, $B$ can be fitted with $B=B_0(L)+B_1(L)/g+B_2(L)/g^2$ with $B_0(L)$ being a linear
function of $(\tilde{L}/\ell_0)^{-1}$. In the limit of large $L/\ell_0$, $B_0$ approaches $0.96$. Thus, our results indicates
that the analytical expression of \cite{Mirlin00} is valid only when both $g$ and $L/\ell_0$ are sufficiently large.

\end{subsection}

\begin{subsection}{(b) crossover from quasi-1D to slab geometry}
It is interesting to see how the behavior of $D(t)$ crosses over from a quasi-1D to a slab geometry as $2R/L$ increases. A plot
of $D(t)/D_0$ as a function of dimensionless time $t^\prime=t/\tau_D$ at $2R/L=0.35$, $0.5$, $1$, $2$, $4$ and $\infty$ (slab)
is shown in Fig.~3 for the case $L=20\ell_0$ and $k\ell_0=4$.  When $2R/L=0.35$, the $\delta_1$ term dominates and $D(t)/D_0$
decreases rapidly between $t\simeq 2 t_D$ and $15t_D$ and gradually saturates to a constant value at later times. As $2R/L$
increases, the peak of $D(t)/D_0$ near $t\simeq 2-3t_D$ also increases. In the meantime, the decay of $D(t)/D_0$ at later times
is reduced.  This indicates a smaller WL effect when $2R/L$ is increased. This is consistent with the decrease of $\delta(0)$
with $2R/L$ shown in the inset of Fig. 1. When $2R/L >2$, $D(t)/D_0$ approaches a constant value.  This again is consistent with
the saturation of $\delta(0)$ when $2R/L>2$ shown in the inset of Fig. 1.

\begin{figure}
\includegraphics [width=\columnwidth] {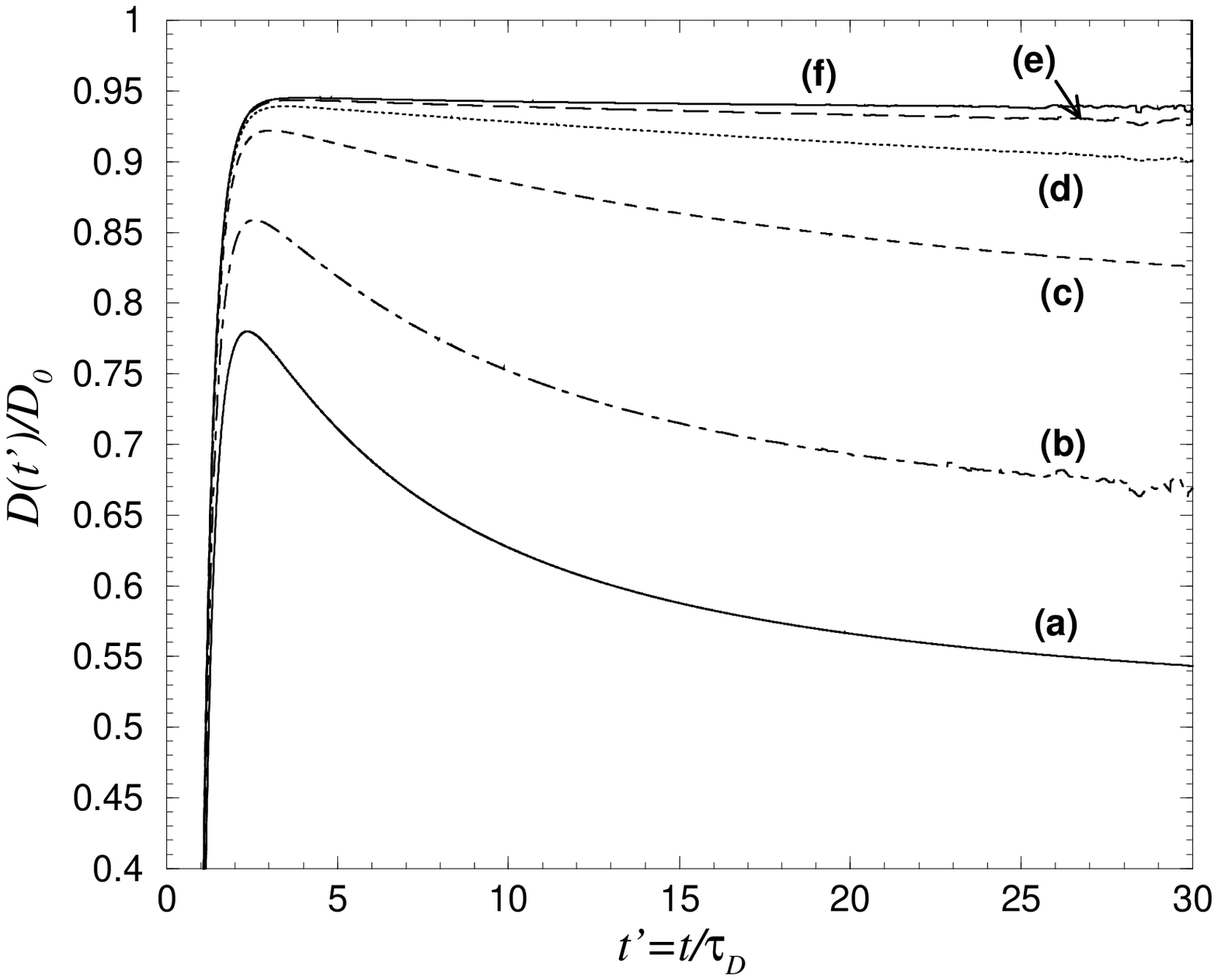}
\caption{$D(t)/D_0$ is plotted as a function of $t^\prime=t/\tau_D$ for the case of $L=20\ell_0$ and $k\ell_0=4$, with (a)
$2R/L=0.35$, (b) $2R/L=0.5$, (c) $2R/L=1$, (d) $2R/L=2$, (e) $2R/L=4$ and (f) $2R/L\rightarrow\infty$ (slab).}
\end{figure}

In order to give a systematic description of the crossover behavior, we present the crossover behavior of the intercept A
obtained from the linear fit of $D(t)/D_0$ immediately after the peak.  In Fig.~4, we plot $A$ as a function of $g$ for
different values of $2R/L$ and for $L/\ell_{0}=20$. The data are well fit by $A=A_{0}+A_{1}/g+A_{2}/g^2$. Thus, at large g, we
have $A=1-C_0(2R/L)/g$, where $C_0$ is an increasing function of $2R/L$. From the fit of the data (dashed lines), we find that
$C_0(2R/L)$ is proportional to $R^2$ when $2R/L>1$. This $R^2$ dependence cancels the $R^2$ factor in $g$ in the denominator.
Thus, in the slab limit, we can write $A=1-c(L)/(k\ell_0)^2$. This behavior can also been seen from the inset of Fig.~4, where
$A$ is plotted as a function of $2R/L$ for different values of $k\ell_0$ for the case of $L=20\ell_0$. $A$ approaches a constant
value when $2R>L$ for each value of $k\ell_0$.

\begin{figure}
\includegraphics [width=\columnwidth] {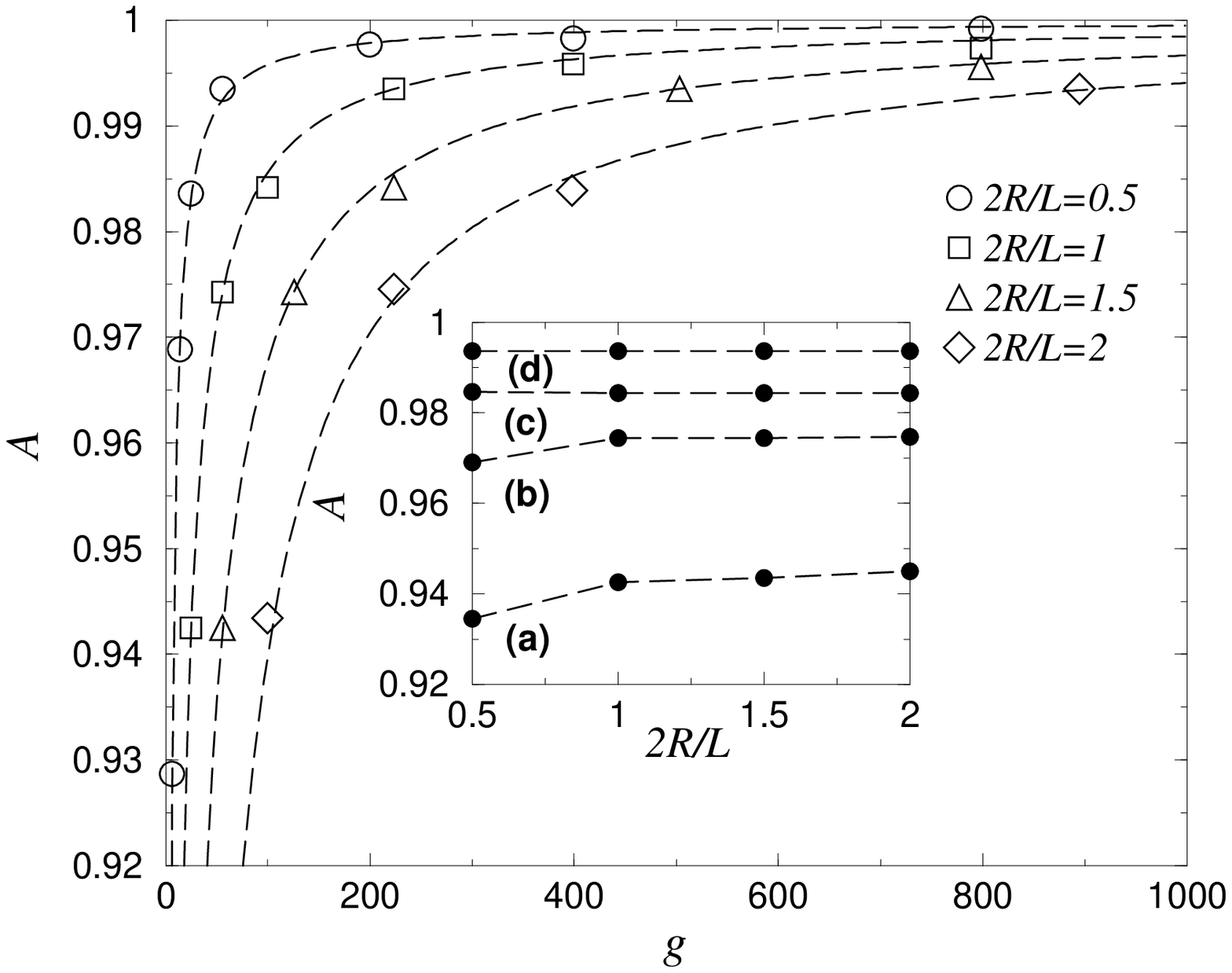}
\caption{The parameter $A$ obtained from the fit of $D(t^{\prime\prime})/D_{0}$ with $A-Bt^{\prime\prime}$ is plotted as a
function of $g$ for $L/\ell_0=20$ and different aspect ratio $2R/L$. The dashed curves are the fit of $A$ with
$A=A_{0}+A_{1}/g+A_{2}/g^2$, where $A_{i}$ are constant. In the inset: $A$ versus $2R/L$ for $k\ell_0=$ (a) $4$, (b) 6, (c) 8
and (d) 12.}
\end{figure}

\end{subsection}

\begin{subsection}{(c) slab geometry}
In order to find the function $c(L)$ in $A(k,L)=1-c(L)/(k\ell_0)^2$ for the slab, we plot $A(k,L)$ as functions $1/(k\ell_0)^2$
in Fig.~5 for different values of $L/\ell_0$.  The slopes of these curves give $c(L)$ and is shown in the inset of Fig. 5. Since
$c(L)$ is an increasing of $L$, $D(t)/D_0\cong A$ decreases with $L$.  This reflects continuous renormalization of the diffusion
coefficient with increasing $L$  due to the presence of longer recurrent scattering path lengths in a thicker sample. The
fitting of the data of $c(L)$ in the inset of Fig.~4 gives $c(L)=1.02-3.72\ell_0/\tilde{L}$. Thus, we find $D(t)/D_0\cong
1-1.02/(k\ell_0)^2+3.72\ell_0/\tilde{L}(k\ell_0)^2$. In the limit of large $L$, we recover the result of weak localization
theory for the bulk, i.e., $D/D_0=1-3\alpha/\pi(k\ell_0)^2$ with $\alpha=1$ \cite{Kirkpatrick85}.

\begin{figure}
\includegraphics [width=\columnwidth] {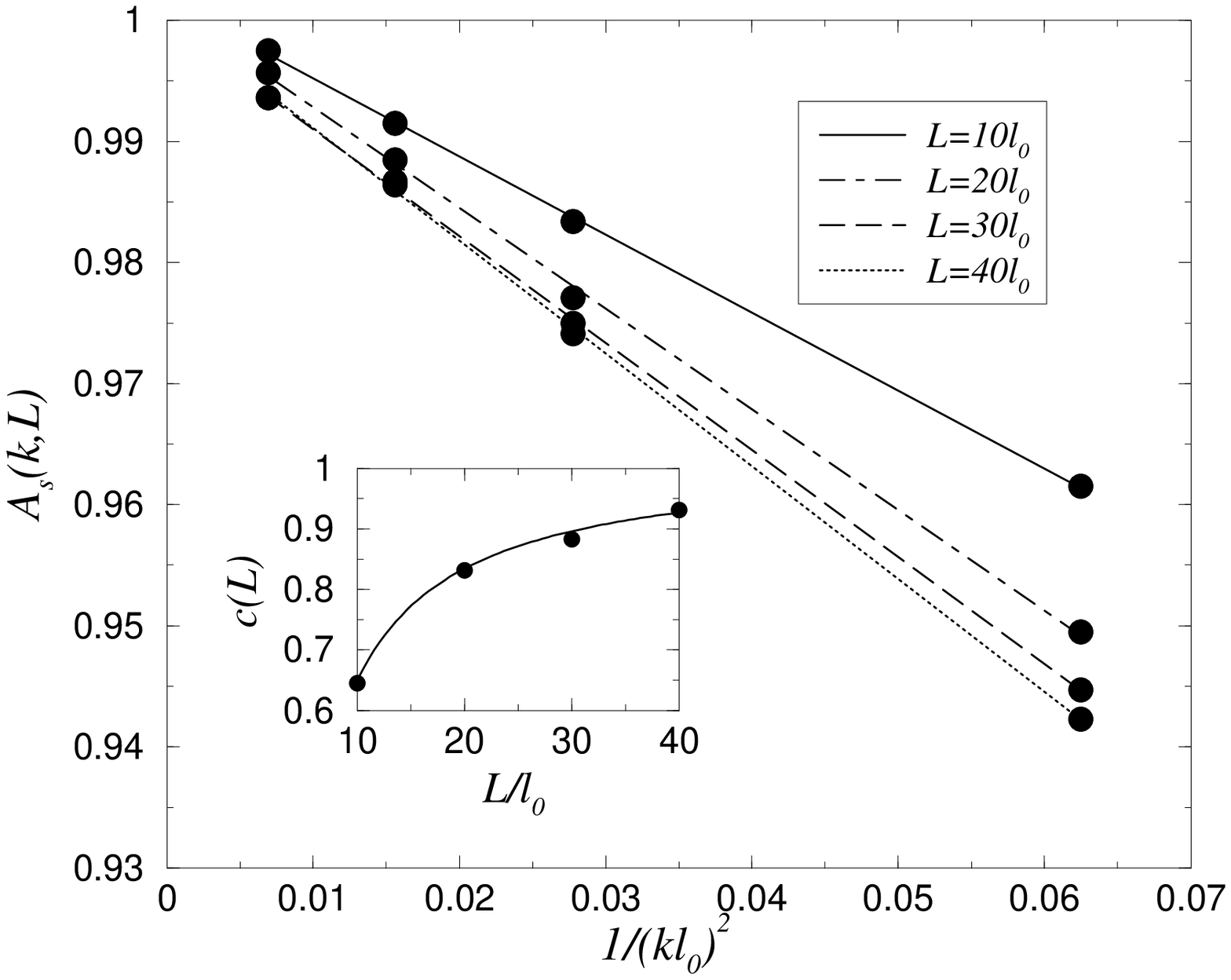}
\caption{The saturation values $A_s(k,L)$ in large $R$ obtained from the inset of Fig. 4 are plotted versus $1/(k\ell_0)^2$ for
the cases of $L=10\ell_0$, $20\ell_0$, $30\ell_0$ and $40\ell_0$. In the inset: $c(L)$ is plotted as a function of $L/\ell_0$,
which is obtained from the fit of the formula $A_s(k,L)=1-c(L)/(k\ell_0)^2$. The solid curve is the fit of the data points with
the formula $c(L)=1.02-3.72\ell_0/\tilde{L}$.}
\end{figure}

\end{subsection}

\end{section}

\begin{section}{IV. Conclusion}
In conclusion, we have solved the Bethe-Salpeter equation with recurrent scattering to find the impact of WL on the dynamics of
wave propagation through a finite-sized mesoscopic sample.  The crossover from quasi-1D to slab geometry is studied through the
change in the behavior of the time-dependent diffusion coefficient $D(t)$ for different aspect ratio $2R/L$. In quasi-1D
systems, $D(t)$ decays rapidly immediately after the peak of the pulse is transmitted. In contrast, $D(t)$ is nearly constant
for a slab. These different behaviors are attributed to two different kinds of WL factors which appear in each case. These
results are consistent with microwave \cite{Chabanov02} and optical \cite{Alfano90,Lagendijk97} measurements.
\end{section} \\

Discussions with P.~Sheng are gratefully acknowledged. This research is supported by Hong Kong RGC Grant No.~HKUST 6163/01P and NSF Grant
No.~DMR0205186.

\end{document}